# Superbunch Hadron Colliders

Ken Takayama, Junichi Kishiro, Makoto Sakuda, and Masayoshi Wake
*High Energy Accelerator Research Organization (KEK)*
Oho, Tsukuba, Ibaraki 305-0801 Japan

A novel concept of a high luminosity hadron collider is proposed. This would be a typical application of an induction synchrotron being newly developed. Extremely long bunches, referred to as superbunches are generated by a multi-bunch stacking method employing barrier buckets at the injection into the collider, and are accelerated to the collider energy with an step-voltage induced in the induction gaps. Superbunches intersect with each other, yielding a luminosity of more than $10^{35}$/cm$^2$/sec. A combination of vertical crossing and horizontal crossing can be used in order to avoid any significant beam-beam tune shift.

PAC numbers: 29.20.Dh, 29.27.Eg

*Introduction*  At CERN, the construction of the Large Hadron Collider (LHC) began in November, 2000, and is scheduled to be completed in 2006. It may be regarded as a second-generation of p-p colliders, succeeding the Intersecting Storage Ring (ISR) at CERN. Recently, serious interest in the Very Large Hadron Collider (VLHC) has rapidly grown in US, expecting to explore new physics beyond the LHC. Both colliders are entirely based on the conventional RF technology for acceleration and the longitudinal confinement of proton bunches. The luminosity is expressed as $L = F \dfrac{k_b N_b^2 f_{rev} \gamma}{4\pi \varepsilon_n \beta^*}$, where $k_b$ is the number of bunches per ring, $N_b$ the number of protons per bunch, $f_{rev}$ the revolution frequency, $\varepsilon_n$ the normalized r.m.s. transverse emittance (assumed to be the same in both planes), $\beta^*$ the betafunction at the interaction point, and $F$, $1/\sqrt{1 + (\Phi \sigma_s / 2\sigma^*)^2}$ ($\sigma_s$, r.m.s. bunch length; $\sigma^*$, r.m.s. beam size at the collision point), the reduction factor caused by the finite crossing angle $\Phi$ [1]. Still, the following conditions must be satisfied: (1) the beam emittance must fit into the small aperture of the superconducting magnets; (2) the total intensity ($k_b N_b$) is limited by the thermal energy produced by synchrotron radiation, which must be absorbed by the cryogenic system; (3) the beam-beam effects cause a spread in betatron tunes and the linear tune shift, $\xi = \dfrac{N_b r_0}{4\pi \varepsilon_n}$ ($r_0$, classical radius of proton), must be kept below a certain limit (0.004); and (4) the longitudinal emittance ($\varepsilon_L$) must be small at injection (small $\Delta p/p$ to ease beam transport from the injector to the collider), and large at collision so as to avoid transverse emittance blow-up by intra-beam scattering.

$k_b$ is mostly determined by bunch-spacing required from the demand of time-resolution of particle detectors; also, the injection kicker rise-time and dump kicker rise-time partially limit the bunch population. Eventually, a significant fraction of possible bunch places (RF harmonics) is vacant. When the beam-occupation ratio against the entire accelerator circumference at injection is defined by $\kappa = \sqrt{2\pi} k_b \sigma_s / C_0$, where $C_0$ is the ring circumference, $\kappa$ is quite small in a conventional hadron collider. It is about 3.4% in the LHC.

If the heat deposited by synchrotron radiation on the cryogenic system can be removed by any efficient engineering efforts and the particle detector doesn't care about the minimum bunch-spacing, the last issue to prevent the collider from reaching a much higher luminosity should be a sparse bunch population that is a limit of the conventional RF synchrotron.   If the proton beams occupy most of the region along the collider circumference with an allowable momentum spread, the luminosity of hadron colliders would drastically increase.   In the proposed scheme, a 70~80% of the circumference can be occupied by the proton beam instead of 3~4%．  This situation is like continuous collisions between proton beams stored in two rings, as seen in Fig.1.

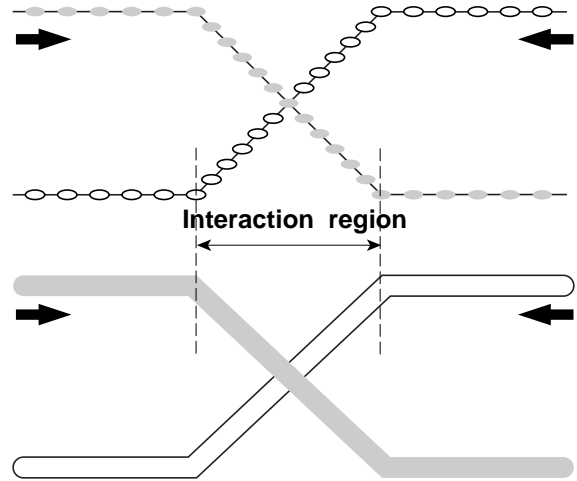

Figure 1 (top). A collision between two bunch trains in the conventional hadron collider; (bottom) a collision between two counter rotating superbunches.

In this letter a novel scheme to realize such continuous collision is proposed.   The luminosity is estimated to be at least twenty-times higher than that in hadron colliders, based on conventional RF technology (hereafter, this type of collider is called the conventional hadron collider, CHC).   The induction synchrotron recently being developed [2] is capable of generating an extremely long microsecond bunch called a superbunch, which keeps the same momentum spread and local intensity as that in the CHC scheme.   After stacking superbunches at the final stage of the collider accelerator complex and accelerating them by a step-voltage generated in the induction gaps to the collider energy, multiple superbunches in both rings are provided for collisions.   The concept of a new type collider, called a superbunch hadron collider (SHC) hereafter, is described together with discussions of beam-physics issues.   Key engineering issues to remove any serious heat-deposit of synchrotron radiation in the SHC are also discussed.   The merit and demerit of the SHC from the standpoint of the detector are discussed.   The application of the induction synchrotron to neutrino physics experiments are also commented.

*Induction synchrotron*   A proton synchrotron employing induction cells (IC) instead of radio-frequency cavities is called an induction synchrotron (IS), and has been described in detail[2]. Acceleration and longitudinal focusing are independently achieved with different induction devices, which consist of an IC loaded by a ferro-magnetic material and a pulse-modulator rapidly switched in synchronization with beam acceleration[3].   As schematically shown in Fig.2, a dc-like induction acceleration is provided by the IC, which is energized with a long voltage-pulse and a short reset-pulse.   The other type of ICs generates a pair of rectangular short-pulses, forming a barrier bucket in the longitudinal phase-space[4]. The rectangular bucket can accommodate particles to its full capacity. The bucket shape tends to create a uniformly diffused longitudinal distribution of the particles apart from both edges of the bucket. This uniformity is important for diminishing the space-charge effects in the transverse and longitudinal directions. This is quite effective to substantially increase the beam intensity in existing synchrotrons [5]. For the same reason, an induction synchrotron is being seriously considered as a possible candidate for future high-intensity hadron accelerators, such as drivers of neutron spallation source [6].

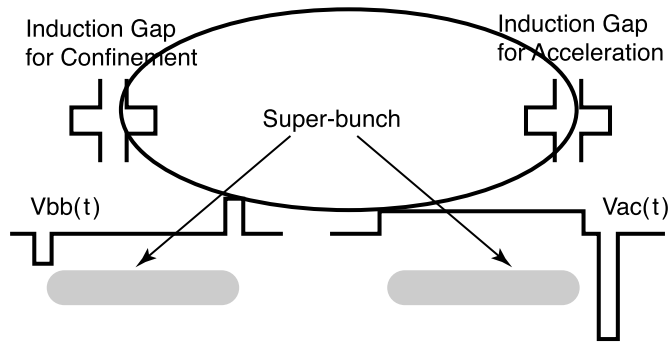

Figure 2. Principle of the Induction Synchrotron.

The ICs are energized with the solid-state power modulator using a fast switching element, such as an array of field effect transistors (FETs) or a Static Inductive (SI) Thyristor, to switch energy from a pre-charged capacitor bank to the IC [7]. Figure 3 shows voltage and core-current pulse-shapes, which were demonstrated in the preliminary low-voltage experiment [3]. The switching frequency of the induction devices corresponding to the revolution frequency is quite important. Rapid excitation generates a serious core loss in the magnetic material of the ICs. Even heat deposited on the switching elements is not ignored. According to the currently achieved engineering capability, it seems to be possible to manage the heat deposit associated with operation at a repetition rate of MHz. Namely, accelerators based on the concept of an induction synchrotron must have a circumference larger than 300m.

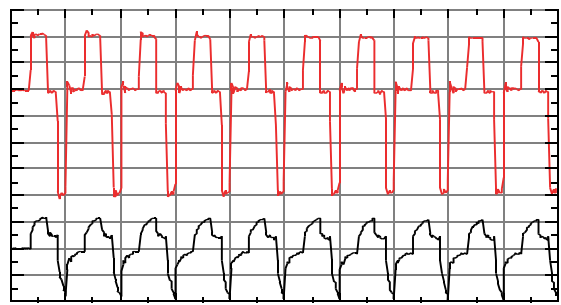

Figure 3. (upper) Pulse shapes of induced voltage (50V/div, 1μsec/div); (lower) core current (10A/div).

The frequency or phase feed-back in RF acceleration, which makes tracking against the ramping magnetic guide-field possible, is replaced by an induction voltage feed-back and a programmable change in the trigger-timing. These feedbacks should be rather simple, unlike that in RF acceleration, where the feedback gain depends on the beam intensity and requires its careful adjustment.

*Multi-bunch stacking by barrier buckets, superbunch formation, and acceleration* An accelerator system for a collider consists of an H$^-$ linac, three-stage booster rings, and a collider ring, as can be seen in the SSC [8] and the LHC [9]. In a case where the circumference of the first booster ring is sufficiently large to employ the principle of IS, a superbunch with a bunching factor of 0.76 can be created by a method called symmetric painting, which is described in detail in references [2] and [6]. Then, multiple superbunches are stacked in the next booster ring, by utilizing two sets of barrier buckets, as shown in Fig.4. A superbunch injected into the next booster ring is captured by a matched barrier bucket. Each superbunch is moved adiabatically toward the edge of the stacking bucket, and is then released into the stacking bucket in such a way that the reset timing for the edge voltage of the stacking bucket is delayed by the bunch-length of the fresh bunch. After this stacking process, a newly generated superbunch is accelerated with the step voltage to the injection energy of the next ring. This process is repeated at every stage of the accelerator rings.

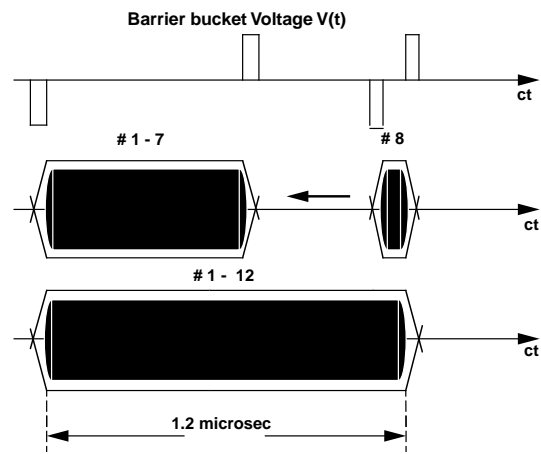

Figure 4. Superbunch formation using two sets of barrier buckets.

*Concept of collider* The maximum pulse-length of the accelerating induction voltage will be limited to order of several microseconds for engineering reasons, such as the practical size of the manufactured induction-core. Thus, the collider must be occupied by multiple superbunches, and bunch spacing is used to reset the magnetic materials. In the case of an LHC-size ring, the number of superbunches is around 40. In principle, the superbunches can occupy a considerable fraction of the ring circumference, 80%, with a momentum spread determined by the barrier-bucket height. After reaching the flat-top energy through collisions, a slight magnitude of accelerating voltage is held to replenish the energy loss due to synchrotron radiation, the superbunches being confined with the barrier buckets. Each of the superbunches intersects with its own counterpart in a half time-period of the bunch length, as shown in Fig.1. The typical parameters required on the ICs employed in the LHC-size SHC are summarized in Table 1, where RF parameters and beam parameters[9] are attached for a comparison. Their operational repetition rate is 1 MHz and the magnitudes of integrated induced voltage for confinement and acceleration are moderate. We will not face any serious engineering difficulties.

Table 1. RF/IC and beam parameters.

|  |  | LHC(Injection/Flat-top) | LHC-size SHC | |
|---|---|---|---|---|
| Total RF/IC voltage | $V_{RF}$ (MV) | 8/16 | $V_{BB}$ (kV) | 25/26 |
|  |  |  | $V_{AC}$ (kV) | 0/(480)/6.7 |
| Bunch area | $A_b$ (eV•s) | 1/2.5 for ($2\sigma_s$) | $5.4\times10^2/1.1\times10^3$ | |
| Bucket area | $A$ (eV•s) | 1.46/8.7 | $7.8\times10^2/1.6\times10^3$ | |
| Bucket half-height | $\Delta p/p$ | $1\times10^{-3}/3.6\times10^{-4}$ | $6\times10^{-4}/1.2\times10^{-4}$ | |
| Bunch length | $\sigma_s$ (m), $\sigma_{sb}$(m) | 0.13/0.075 | 462/267 | |
| Energy spread | $\Delta E/E$ | $4.5\times10^{-4}/1\times10^{-4}$ (rms) | $9\times10^{-4}/1.8\times10^{-4}$ (full) | |
| Bunch numbers per ring | $k_b$ | 2835 | 40 | |

(the parenthesized number in the 4th row is a magnitude during acceleration.)

The luminosity in the SHC, normalized by that in the CHC with the same local beam density and zero crossing-angle, is written in terms of a function of crossing angle $\Phi$ and an effective size ($2l$) of the particle detector [10],

$$\frac{L_{SHC}(\Phi,l)}{L_{CHC}(0)} = 4\frac{(k_{sb}\sigma_{sb})}{(k_b\sigma_s^{'})}\frac{1}{\sigma_s^{'}}\int_0^l \frac{\exp\left[-\frac{\gamma\Phi^2 s^2}{2\beta^*\varepsilon_n(1+s^2/(\beta^*)^2)}\right]}{1+s^2/(\beta^*)^2}ds \ . \qquad (1)$$

For the simplicity, the Gaussian distribution in the longitudinal direction for the CHC is replaced by a rectangular distribution with a bunch length, $\sigma_s^{'} = \sqrt{2\pi}\sigma_s$ and the number of protons per bunch, $N_b$, leading a uniform/peak line density, $\lambda = N_b/\sigma_s^{'}$. Deriving the above equation, similar optics and the Gaussian distribution for both transverse directions are assumed. In the limit of $\Phi = 0$, $k_{sb} = k_b$, $\sigma_{sb} = \sigma_s^{'}$, $2l = \sigma_s^{'}/2$, Eq.(1) becomes unity. The factor of $(k_{sb}\sigma_{sb})/(k_b\sigma_s^{'})$ in Eq(1) represents the relative ratio of beam occupation in the SHC and CHC. As mentioned earlier, the parameter is around a factor of 20. For a typical example, where $\sigma_s^{'} = 15cm$, $\beta^* = 0.5m$, $\varepsilon_n = 1\mu rad$, $mc^2\gamma = 20TeV$, $2l = 5m$, the normalized luminosity is shown as a function of $\Phi$ in Fig.5.

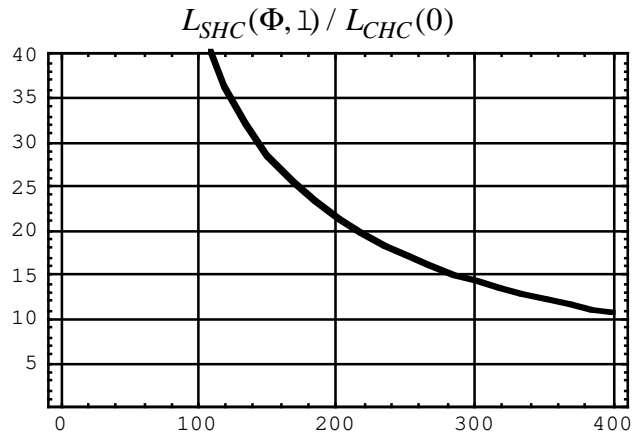

Φ (μrad)

Figure 5. Normalized luminosity.

Rapid decreasing with the crossing angle is remarkable. This comes from a simple reason, that collisions between continuous beams are likely to be affected by the collision angle. In the SHC scheme, the incoherent beam-beam tune shift is of big concern, even with a deep crossing angle. The incoherent beam-beam tune-shift can analytically be evaluated by manipulating the non-oscillating terms in the beam-beam perturbing potential. The tune-shift normalized by that in the head-on collision of the CHC scheme is given in the following forms [10]:

$$\frac{(\Delta v_x)_\Phi^{SHC}}{\xi} = \frac{8\beta^* \varepsilon_n}{\sigma_s' \gamma} \int_0^{l_{int}} \frac{1+s^2/(\beta^*)^2}{\Phi^2 s^2}\left[1-\exp\left(-\frac{\gamma \Phi^2 s^2}{2\varepsilon_n \beta^*\left(1+s^2/(\beta^*)^2\right)}\right)\right]ds, \qquad (2)$$

$$\frac{(\Delta v_y)_\Phi^{SHC}}{\xi} = \frac{8}{\sigma_s'} \int_0^{l_{int}} \exp\left(-\frac{\gamma \Phi^2 s^2}{2\varepsilon_n \beta^*\left(1+s^2/(\beta^*)^2\right)}\right)ds - \frac{(\Delta v_x)_\Phi^{SHC}}{\xi}, \qquad (3)$$

where crossing in the vertical direction is assumed and $2l_{int}$ is the size of the interaction region, $2l \langle\langle 2l_{int} \langle\langle \sigma_{sb}$. In the limit of $\Phi = 0$, $2l_{int} = \sigma_s'/2$, Eqs. (2), (3) become unity. The numerical values for both directions are shown as functions of the crossing angle $\Phi$ in Fig. 6.

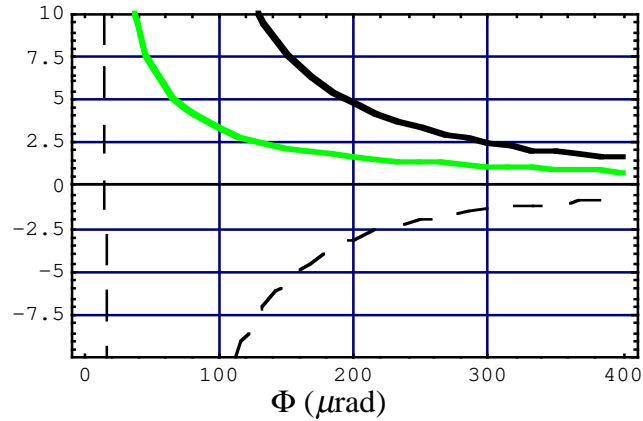

Figure 6 Normalized beam-beam tune shifts for $2l_{int} = 50m$
(solid, horizontal; broken, vertical; gray, sum).

A change in the polarity for the vertical direction beyond some critical crossing-angle is notable. This is understandable from speculating that a particle is focused by space-charge effects as it leaves from the beam-core region, while it is defocused in the core region. Namely, a longer stay outside the core region gives net focusing through the interaction region beyond a certain critical crossing angle. The characteristics strongly suggests that hybrid crossing (vertical crossing in one interaction region (IR) and horizontal crossing in the other) should be employed in the SHC scheme. The collider rings necessarily have twists. By hybrid crossing, as schematically shown in Fig. 7, the beam-beam tune-shift largely diminishes for both directions. The relative tune-shifts for both directions are less than 2.0 for the crossing-angle beyond 150μrad, where the luminosity is quite attractive, as found in Fig.5. The magnitude is sufficiently acceptable, because it is equal to the integrated head-on beam-beam tune-shift in the CHC scheme with a couple of IRs.

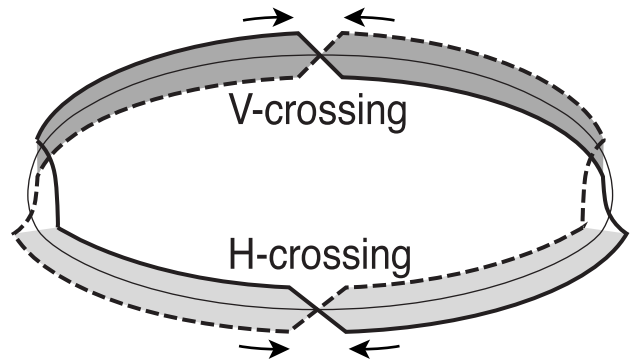

Figure 7. Schematic view of hybrid crossing.

*Beam physics issues:* Long-range beam-beam interaction, intra-beam scattering, e-p instability
Beam-beam effects are not regarded as being a serious limitation of the SHC. The beam-beam tune-shift parameters during nominal operations are rather similar to that in the CHC. However, the beam-induced orbit distortions should be much larger, because two superbunches travel close to each other in the same pipe. As discussed above, any long-range beam-beam effects can be reduced by using a larger crossing angle, but then the nonlinear fields in the IR triplets are enhanced because the beam goes through them more off-center. Long term effects are now being investigated by utilizing a very fast computer simulation code, PATRASH, which was originally developed to study halo-formation in high-intensity proton rings [11]. The impact of intra-beam scattering on the luminosity lifetime depends on the collider parameters. A feature of intra-beam scattering in the SHC is analogous to that of the coasting beam. At least the intra-beam scattering growth time in the longitudinal direction should become remarkably long; the growth time is dominated by that in the transverse direction.

*Heat deposit due to synchrotron radiation and Joule loss due to wall current* Synchrotron radiation has a significant impact on the collider design because (1) the radiation power must be absorbed at liquid-helium temperature; (2) photons desorb molecules from the vacuum chamber wall and accumulated photo-electrons may become a resource of the e-p instability; and (3) the radiation damping time affects the luminosity lifetime. The effect of synchrotron radiation on the cryogenic system largely depends on the total radiated power. Since the beam occupation ratio is proportional to the number of stored protons, the radiation power is larger by a factor of 20 than that in the CHC. On the other hand, Joule loss due to the wall current, in which high frequency components in the wall-current spectrum are known to share most of the fraction, is of almost same order as that in the CHC, because the bunch spacing of the superbunch train is extremely large. The magnitude of the loss per unit-length is negligibly small compared with the synchrotron radiation; 47mW/m for the LHC [12] in the CHC scheme (assumed a Gaussian distribution in time) and 112mW/m for the LHC in the SHC scheme (assumed a rectangular shape). The synchrotron radiation heat must be removed primarily by water cooling at room temperature. Interception between room temperature and liquid helium temperature will be cooled by helium gas to 30K. Such a dual radiation shield requires a slightly more complicated structure, but can be accommodated in the beam tube.

*Physics impact* From a physics point of view, the luminosity is essential at any collider to search for new particles, such as Higgs and Super-Symmetric particles[13]. A possible disadvantage of this scheme is the longer collision area along the beam and the overlap of events in a superbunch. However, if we consider a reasonable crossing angle, 200 μrad for LHC, we can design the collision area to be well inside the vertex detector, which covers about 1m in the beam direction. We usually have subdetectors with good timing information (nanoseconds) to resolve the overlap of multi-events within a bunch. Thus, any difficulty with a superbunch will be overcome as long as the rate for events of interest is not very high. In addition, we note that since the local particle density is uniform over the bunch, the event rate per unit time is uniform with the SHC scheme and is even smaller than that with the CHC when the total number of particles is the same.

Meanwhile, if the superbunch scheme is employed in the upstream boosters, it should have a large impact on neutrino oscillation experiments, such as K2K [14], MINOS [15] and OPERA/ICANOE [16], where a long bunch structure of the order of μsec produces no problem and the intensity increase by a factor or an order of magnitude is crucial when the oscillation parameter $\Delta m^2$ is smaller than $3 \times 10^{-3} (eV)^2$.

*Summary* A novel hadron collider based on the induction synchrotron (Superbunch Hadron Collider) has been proposed. Collisions between μsec-long bunches (superbunch) have been shown to give an extremely high luminosity, that is, $10^{35}$/cm$^2$/sec, assuming that the total number of protons is larger by a factor of 20 than that in the CHC. In order to decrease an unavoidable large incoherent beam-beam tune shift, a hybrid collision where one collision takes place in the vertical direction and the other in the horizontal direction is required. Since synchrotron radiation is huge, the deposited heat may be shielded from the liquid-helium temperature outer vacuum-wall by water-cooling the beam-screen. Finally, we emphasize that the present SHC is sufficiently worth to be considered as a possible scheme for the coming generation of hadron colliders such as the VLHC.

The authors acknowledge Takahiko Kondo and Susumu Igarashi for comments on the detector.

*References*
[1] M. Sands, The Physics of Electron Storage Rings, in Proc. of Int. School of Physics "Enrico Fermi", Course 46 "Physics with Intersecting Storage Ring", edited by B.Touschek, (Academic Press, 1971).
[2] K.Takayama and J.Kishiro, "Induction Synchrotron", *Nucl. Inst. Meth.*, **A451/1**, 304 (2000).
[3] J.Kishiro, K.Takayama, E.Nakamura, K.Horioka, M.Watanabe, A.Tokuchi, and S.Naitoh, *Proc. of EPAC2000*, 1966-1968 (2000).
[4] An alternative to generate the barrier bucket may be a combination of broad-band low-Q cavity and wide-band amplifier proposed in the reference (Fermilab Recycler Ring: Technical Design Report, November 1996).
[5] K.Takayama, in Neutrino Oscillations and their Origin, Frontiers Sciences No.35, 115 (2000), Universal Academy Press, Tokyo
[6] K.Takayama, *Proc. of ICANS-XV (2000)* in press
[7] J.Kishiro, *Proc. of ICANS-XV (2000)* in press
[8] Superconducting Super Collider Laboratory, Site-Specific Conceptual Design, July 1990.
[9] P.Lefevre, T.Petterson (Ed.), The Large Hadron Collider, Conceptual Design, CERN/AC/95-05(LHC), Oct. 1995.
[10] The evaluated formula is in agreement with the result given by E.Keil, C.Pellegrini, and A.M.Sessler (*Nucl. Inst. Meth.*, **113**, 333 (1973) and CRISP Report 72-34 (BNL), 1972).
[11] Y.Shimosaki and K.Takayama, submitted to *Phys. Rev. E.*
[12] Our calculation gives this number, while the reference [9] claims 75 mW/m.
[13] H.G.Evans, hex-ex/0007024.
[14] K.Nishikawa, *Nucl. Phys.B* (Proc. Suppl.)**77**, 198(1999).
[15] S.G.Wojcicki, *ibid.,* 182 (1999).
[16] P. Picci and F. Pietropaolo, *ibid.*, 187 (1999).